\documentstyle[preprint,revtex]{aps}
\tightenlines
\draft
\begin{document}

\begin{title}
A Non-Standard Quark Model and the Mass Difference $\Xi^- - \Xi^0$
\end{title}
\author{J. Tinka Gammel$^{a}$ and John L. Gammel$^{b}$}
\begin{instit}
$^{a}$ Code 573, Materials Research Branch, \\
NCCOSC RDT\&E Division, San Diego, CA 92152-5000 \\
$^{b}$ Professor of Physics Emeritus, \\
Saint Louis University, St. Louis, MO 63103
\end{instit}
\begin{abstract}
We show that a highly unconventional theory, based on the
16 dimensional (${3\over2}$,${3\over2}$) irreducible representation of O(4),
fits the masses of the baryons
$\Xi^-$,$\Xi^0$,$\Sigma^-$,$\Sigma^0$,$\Sigma^+$,$\Lambda$,$p$,$n$ and
the leptons $\mu$,$e$,$\nu_\mu$,$\nu_e$ to within current experimental
error, yet yields
a prediction substantially different from that of Coleman and
Glashow, namely $\Xi^- - \Xi^0 < 6.1$ MeV. We discuss the
ramifications of this very non-standard model in the hope to
stimulate improved measurements of the $\Xi^- - \Xi^0$ mass
difference.
\end{abstract}

\begin{narrowtext}
\narrowtext

\section{Introduction}

     The standard model \cite{r3} in elementary particle physics is well
established. For instance, the Coleman-Glashow \cite{r4} mass formula
\begin{equation}
\Xi^- - \Xi^0 = \Sigma^- - \Sigma^+ - (n - p),
\end{equation}
predicts \cite{r5}
\begin{equation}
\Xi^- - \Xi^0 = 6.78 \pm 0.08 {\rm ~MeV}.
\end{equation}
The experimental value is
\begin{equation}
\Xi^- - \Xi^0 = 6.4 \pm 0.6 {\rm ~MeV}.
\end{equation}

Close \cite{r4} says the prediction is ``perfect''. In
view of the large experimental error in the measurement of
$\Xi^- - \Xi^0$ it is.  It must be difficult to improve the accuracy since
31 years have elapsed since Coleman and Glashow made their
prediction.

     While it is also difficult to take a prediction not based on
the standard model seriously, it may be that a prediction different
from that of Coleman and Glashow will stimulate some
experimentalist to do the impossible and improve the accuracy  of
the experimental value of $\Xi^- - \Xi^0$. We put forward a theory which
contains almost, or quite, unacceptably unconventional features and
which has only the virtue that it produces a definite and in
principle testable prediction different from that of Coleman and
Glashow.  Our prediction is
\begin{equation}
\Xi^- - \Xi^0 < 6.1 {\rm ~MeV}.
\end{equation}

\section{The Theory}

     The mass of the muon is comparable to the mass differences
$\Xi - \Sigma$, $\Sigma - \Lambda$, and $\Lambda - p$.
Also, the mass of the electron is
comparable to the ``electromagnetic'' mass differences such as $n-p$.
So, although it is far fetched to do so \cite{r0},
we postulate that there exist eight otherwise
unknown baryons such that their mass
differences are exactly lepton masses,
\begin{eqnarray}
B_\mu - B_\mu^\prime &=& \mu,  \nonumber\\
B_e - B_e^\prime &=& e,                 \\
B_1 - B_1^\prime &=& \nu_\mu,   \nonumber\\
B_2 - B_2^\prime &=& \nu_e.    \nonumber
\end{eqnarray}

Thus there are 16 baryons (the standard set
$\Xi^-$,$\Xi^0$,$\Sigma^-$,$\Sigma^0$,$\Sigma^+$,$\Lambda$,$p$,$n$ and
eight non-standard baryons
$B_\mu$,$B_\mu^\prime$,$B_e$,$B_e^\prime$,$B_1$,%
$B_1^\prime$,$B_2$,$B_2^\prime$).

     Next, we postulate that the wave functions for these baryons
are a basis for the 16-dimensional (${3\over2}$,${3\over2}$) irreducible
representation of O(4). The rest of the paper shows how masses
are calculated from such a postulate,
explains what this postulate means,
and reports the results of our calculations.

\section{An Example of Mass Calculations}

     Suppose there are two particles $m_1$ and $m_2$ and that their wave
functions are $\alpha$ and $\beta$, the bases for a spin
or iso-spin $S$=${1\over2}$
representation of O(3),respectively.  Then, since the mass terms in
the Lagrangean are quadratic in the wave functions, the direct
products
$\alpha$$\alpha$,$\alpha$$\beta$,$\beta$$\alpha$,$\beta$$\beta$
might appear in the Lagrangean, but one
supposes that these should be linearly combined to form bases for
irreducible representations of O(3).  The coefficients in the
linear combinations are Clebsch-Gordon coefficients.  In the
present case, the linear combinations are bases for the $S$=1 and $S$=0
irreducible representations of O(3), namely
\begin{eqnarray}
                   &   S=1           &~~~~~~~~~  S=0 \nonumber\\
         1~~~   &   \alpha\alpha                       \\
   S_z = 0~~~  & (\alpha\beta+\beta\alpha)/\sqrt{2}
                             &~~~~  (\alpha\beta-\beta\alpha)/\sqrt{2}
                                                         \nonumber \\
        -1~~~   &   \beta\beta                          \nonumber
\end{eqnarray}
{}From the arguments that only $S_z$=0 terms should be
included in the Lagrangean and that $S$=0 should lead to
the simplest mass terms of
all (namely, all masses equal), one is led to

\begin{equation}
         \alpha = \bar\beta , ~~~~~~\beta = -\bar\alpha ,
\end{equation}
so that the $S_z$=0 terms become
\begin{eqnarray}
      &  S=1            &~~~~~~~~~~   S=0   \nonumber\\
S_z=0~~~ &-\bar\alpha \alpha + \bar\beta \beta
                   &~~~~   \bar\alpha \alpha + \bar\beta\beta ,
\end{eqnarray}
where we omit the now unnecessary normalizing factor
$\sqrt{2}$ (it can be absorbed in the c's in
Eq.(\ref{E9}) below) in order to be dealing
with integers.  We call $-\bar\alpha\alpha + \bar\beta\beta$
and $\bar\alpha\alpha + \bar\beta\beta$ ``mass forms''.

    The mass term in the Lagrangean is a linear combination of
the mass forms,
\begin{eqnarray}
   {\cal L} &=& c_1(\bar\alpha\alpha + \bar\beta\beta)
           + c_2(-\bar\alpha\alpha + \bar\beta\beta), \label{E9}\\
     &=& m_1 \bar\alpha\alpha + m_2\bar\beta\beta.   \nonumber
\end{eqnarray}
In general,
\begin{equation}
       m_i = v_{ij}c_j,
\end{equation}
where, in the present example, the matrix of coefficients $v_{ij}$ is
\begin{equation}
({\bf v}) =   \left(
\begin{array}{rr}
    1&  -1  \\
    1&   1 \\
\end{array}      \right)
\label{E11}
\end{equation}
which is such that the columns are a complete set of orthogonal
vectors. In the Appendix, we construct the analog of Eq.(\ref{E11}) for
the (${3\over2}$,${3\over2}$) representation of O(4).
The columns of the 16$\times$16 matrix (${\bf v}$),
shown in Table~\ref{T3},
are a complete set of orthogonal vectors. Nothing has been done
other than to transform the mass data $m_i$ into
mass data $c_i$.  Referring again to the simple example,
if $c_2$=0 then $m_1$=$m_2$.  One regards $c_2$=0 as a simpler
statement than $m_1$=$m_2$ because in the 16-dimensional case a
statement that $c_i$=0 for some particular value of $i$ is a linear
relation among the 16 $m_i$ which is much more complicated than
$m_1$=$m_2$ (as may be seen by looking at Eq.(\ref{E20})).

     Since the columns of (${\bf v}$) form a complete set of orthogonal
vectors which could be made orthonormal (by not dropping the
normalizing factors such as $\sqrt{2}$), (${\bf v}$) essentially
rotates the mass vector ($m_1$,$m_2$,...,$m_{16}$)
into the vector ($c_1$, $c_2$,...,$c_{16}$ ).  A mass
relation $c_i$=0 is a statement that the $c$-vector lacks an $i$-th
component.  This scheme for discovering mass relations, that is,
relations among the $m_i$ in a quadratic form by rotating the vector
($m_1$,$m_2$,...,$m_{16}$) does not need to be defended by physical
arguments. The scheme is a mathematical method.  The physics
(such as it is) lies in the selection of the rotations to be
considered, that is, in the selection of the 16-dimensional
(${3\over2}$,${3\over2}$) irreducible representation of the
group O(4).

\section{Reasons for Working with O(4)}

     One reason for choosing to work with O(4) is technical: 
O(4) is simply related to O(3).  The irreducible
representations of O(3) and the Clebsch-Gordon coefficients for
O(3) are familiar in physics.  The connection between O(4) and
O(3) is described in detail in the Appendix. Another reason is
philosophical.  O(4) is closely related to the Lorentz group.  If
one accepts that the properties of matter and the properties of
space-time must somehow be similar, or even identical, then since
the properties of space-time are governed by the Lorentz group,
it is natural to suppose that the properties of matter are
also governed by the Lorentz group \cite{r0}.
The last reason is that once one has chosen to work with the
(${3\over2}$,${3\over2}$) representation of O(4), then there exists a
connection between the fact that the eight standard baryons form
a singlet ($\Lambda$), two doublets (($\Xi^-$,$\Xi^0$),($p$,$n$)),
and one triplet ($\Sigma^-$,$\Sigma^0$,$\Sigma^+$) and the fact that
three leptons ($\nu_e$,$\nu_\mu$,$e$) have small ``electromagnetic''
masses while one ($\mu$) has a large mass. We explain this connection now.

     We have postulated that the wave functions of the 16 baryons
are a basis for the 16-dimensional (${3\over2}$,${3\over2}$) representation
of O(4).  The finite dimensional irreducible representations of O(4)
(see the Appendix) are all characterized by
two spins or iso-spins $S_1$ and $S_2$,
denoted by ($S_1$,$S_2$).  The wave functions which form a basis for
such a representation are distinguished (that is, labeled)
by two quantum numbers $S_{1z}$ and $S_{2z}$
where $-S_1$$\le$$S_{1z}$$\le$$S_1$ and $-S_2$$\le$$S_{2z}$$\le$$S_2$.
The dimensionality of the representation is (2$S_1$+1)(2$S_2$+1). Thus
$S_1$=${3\over2}$,$S_2$=${3\over2}$ suggests that the baryons form a quartet
of quartets as shown in Table~\ref{T1}.  We assign the cascades to
the top quartet, the sigmas to the second quartet, the lambda to
the third quartet, and the proton and neutron to the bottom
quartet. One $B$ particle must be placed in the sigma group in
order to complete the sigma group.  The corresponding $B^\prime$ must be
placed in another group, so that $B-B^\prime$ is of order 100 MeV.  It is
obvious that these $B$'s are the $B_\mu$ and $B_\mu^\prime$.
$B_\mu^\prime$ may be placed in
the lambda group so that the remaining pairs
(($B_e$,$B_e^\prime$),($B_1$,$B_1^\prime$),($B_2$,$B_2^\prime$))
may each be placed in the same group
so that the mass differences are at most ``electromagnetic''.
This explains the connection between the baryon multiplets and
lepton mass groupings as stated above.

     Thus, in the manner just described, we assign each of the 16
baryons to one of the 16 states in Table~\ref{T1}.  The total number of
possible assignments is \hfil\break
\begin{tabular}{lp{5.2in}}
3$\times$2
       & (3 is the number of ways of choosing which of the pairs
          ($B_e$,$B_e^\prime$), ($B_1$,$B_1^\prime$),
          ($B_2$,$B_2^\prime$) is to be placed in the cascade
          group and 2 is the number of ways of choosing which of
          the remaining pairs is to be placed in the lambda
          group.) \\
$\times$(4!)$^4$
       & (4! is the number of permutations of the 4 particles
          within each group, and this must be raised to the 4th
          power because there are 4 groups.) \\
{}/2$^3$
       & (because, in mass calculations,
          $B_1$,$B_1^\prime$,$B_2$,$B_2^\prime$ are all
          indistinguishable) \\
\end{tabular}
=248,832.

It is possible to sort through all possible assignments and
find the one which best fits the experimental data.
However, to reduce the degree of arbitrariness and increase
the physical significance of the work \cite{r0},
we now introduce further structure.

\section{Quark Model}

     In order to reduce this large number of possible assignments,
we allow only those assignments which agree with a ``quark'' model
(not related to the standard quark model, although in some cases
there is a striking similarity -- the fractional 1/3 and 2/3
charges which appear in some cases, for example).

     (${3\over2}$,${3\over2}$) suggests that each baryon
is made out of three (${1\over2}$,${1\over2}$) ``quarks'';
that is, there are four quarks which we call
$\alpha^\prime\alpha$,\,$\alpha^\prime\beta$,\,%
$\beta^\prime\alpha$,\,$\beta^\prime\beta$
($\alpha^\prime,$\,$\beta^\prime$ are bases for
a spin or iso-spin S=${1\over2}$
representation of O(3), as are $\alpha,\beta$).
If $q_1,q_2,q_3,q_4$ are the electric charges of
$\alpha^\prime\alpha$,\,$\alpha^\prime\beta$,\,%
$\beta^\prime\alpha$,\,$\beta^\prime\beta$,
respectively, then the baryon
charges are as shown in Table~\ref{T1}. If we suppose that the baryons
have integer charges, then we find that only the six
possibilities listed in Table~\ref{T2} exist. The total number of
assignments is \hfil\break
\begin{tabular}{lp{5.2in}}
6         & (the six cases just described)  \\
$\times$3 & (the number of ways of placing $B_\mu^\prime$ in the lambda
group)\\
$\times$6 & (because the ($B_e$,$B_e^\prime$) can be placed in the cascade,
             lambda,or nucleon group in two orders)  \\
$\times$1 & (there are no more factors because
              $B_1$,$B_1^\prime$,$B_2$,$B_2^\prime$ are
             indistinguishable in mass calculations)  \\
\end{tabular}
=108, a considerable reduction.

\section{Mass Relations}

     One may write 12 equations for the experimentally known
masses \cite{r6} (call them $m^\prime$) $\Sigma^-$, $\Xi^- - \Sigma^-$,
$\Sigma^- - \Lambda$, $\Sigma^- - p$, $\Xi^- - \Xi^0$,
$\Sigma^- - \Sigma^0$, $\Sigma^- - \Sigma^+$, $p - n$,
$\Lambda + \mu - \Sigma^-$, $e$, $\nu_\mu$, and  $\nu_e$ in terms of
the 16 coefficients $c_i$ of the 16 mass forms given in the Appendix for O(4),
\begin{equation}
m_i^\prime = d_{ij} c_j , ~~ i=1 {\rm ~to~}  12, ~~j=1 {\rm ~to~} 16.
\label{E12}
\end{equation}
The $d_{ij}$ involve differences of the $v_{ij}$ given in the Appendix.  It
is not possible to solve Eq.(\ref{E12}) for the $c_j$ unless four of the $c_j$
are chosen to be zero.  Then
\begin{equation}
    c_i = (d^{-1})_{ij}m_j^\prime ,~~ $j$=1 {\rm ~to~} 12,
\end{equation}
where $i$ ranges over the 12 values of $i$ such that $c_i$ has not
been chosen to be zero.
If, in addition to the four $c_i$ chosen to be zero to make Eq.(\ref{E12})
soluble, another $c_i$ is chosen to be zero (call it $c_p$), then
\begin{equation}
              c_p = 0 = (d^{-1})_{pj}m_j^\prime
\end{equation}
is a linear mass relation among the twelve $m_j^\prime$.

     $c_1$,$c_2$,$c_3$, and $c_4$ determine the average value of the masses
of the  particles within each of the cascade,sigma,lambda, and
nucleon groups, and assuming (as we do) that the $B$'s do not have
masses much different from the masses of the other particles in
the group to which they belong, none of the first four $c$'s may be
expected to be zero.  $c_5$ through $c_{16}$ determine the
``electromagnetic'' splittings.  We choose six of these to be zero
so that we predict 6-4=2 ``electromagnetic'' mass relations. We
examine all possible ways of choosing six of $c_5$ through $c_{16}$ to be
zero, and all possible pairs of mass relations which result.
There are 12!/6!6!=924 ways in which six of the twelve $c_5$
through $c_{16}$ can be chosen to be zero; thus, the number of pairs
of mass relations which we examine is 108$\times$924=99,792.

     We use the mass relations to calculate the masses $\Xi^- - \Xi^0$
and $\Sigma^- - \Sigma^0$ in terms of the other ten experimentally known
masses,
\begin{eqnarray}
       \Xi^- - \Xi^0 &=& a_j m_j' ,  \\
       \Sigma^- - \Sigma^0 &=& b_j m_j' ,  \nonumber
\end{eqnarray}
where $m_j'$ is not  $\Xi^- - \Xi^0$ or $\Sigma^- - \Sigma^0$.
We reject predictions involving large cancellations by
requiring \cite{r7}
\begin{equation}
           (\Xi^- - \Xi^0)/(|a_j ||m_j'|) > 1/3
\end{equation}
and similarly for $\Sigma^- - \Sigma^0$.

     We summarize the results in the graphs of $\Sigma^- - \Sigma^0$ versus
$\Xi^- - \Xi^0$.  In Fig.~\ref{fig1}, the 99,792 pairs of mass predictions
have been ``binned'' into 200x200 pixels spanning the range
($-$40:40,$-$40:40). The size of the square plotted at each pixel is
proportional to the logarithm of the number of mass predictions
within the bin at that point. Clearly, the values predicted most
often do not lie near the experimental data. In Fig.~\ref{fig2}, each pair
of mass predictions is a point on the graph. The point has a size
(or scatter) determined by the experimental errors in the ten
experimental masses from which the point was calculated.  The
scatter was calculated by randomly varying these ten masses
(within limits determined by the experimental errors) and
repeating the calculations.

     In view of the large number of points and the wide range in
the predictions, it cannot be considered worthy of note \cite{r8} that one
of them agrees with the experimental data. However, from the
figures it is clear that one can make a quite general statement:
for all points  such that the predicted value of $\Sigma^- - \Sigma^0$ is
within 1 MeV of the experimental value and $\Xi^- - \Xi^0$ is less than
10 MeV,
\begin{equation}
              \Xi^- - \Xi^0 < 6.1 {\rm ~MeV.}  \nonumber
\end{equation}
This prediction is quite different from the Coleman-Glashow
prediction, but experiment (while agreeing with both predictions)
cannot distinguish the two at present.

     As we said before, we do not regard it as especially
noteworthy that there exist points which lie close to the
experimental data. Still, their properties may be of interest.
The point which lies closest to the experimental
data corresponds
to case II in Table~\ref{T2} and has the following assignment of
particles to the states 1-16 shown in Tables~\ref{T1}~and~\ref{T2}, in order
1-16,
\begin{eqnarray}
\Xi^0, \Xi^-, B_e^\prime, B_e,~~ && \Sigma^+, \Sigma^0, \Sigma^-, B_\mu,
\label{E18}\\
  ~~B_1, B_1^\prime, \Lambda, B_\mu^\prime, && ~~B_2, B_2^\prime, p, n,
\nonumber
\end{eqnarray}
and the $c$'s which are zero are
\begin{equation}
    c_7 = c_8 = c_9 = c_{10} = c_{13} = c_{16} = 0.
\end{equation}
The mass predictions are
\begin{eqnarray}
\Xi^- - \Xi^0& =& (5/4)e + (3/8)(\Lambda + \mu - \Sigma^- )
\nonumber\\
              &&  - (9/8)(n - p) - (1/4)(\Sigma^- - \Sigma^+),
\label{E20}\\
\Sigma^- - \Sigma^0 &=& (3/4)e + (1/8)(\Lambda + \mu - \Sigma^- )
\nonumber\\
              &&  - (3/8)(n - p) + (1/4)(\Sigma^- - \Sigma^+).
\nonumber
\end{eqnarray}
The non-zero $c$'s are (in MeV)
\begin{eqnarray}
c_1 &=1146.7,    ~~~~c_6    &=-1.072,  \nonumber\\
c_2 &=~~~62.11,   ~~~~c_{11} &=-0.562,  \nonumber\\
c_3 &=~-10.48,   ~~~~c_{12} &=-0.886,  \nonumber\\
c_4 &=~~~~7.31,   ~~~~c_{14} &=~~0.421,  \nonumber\\
c_5 &=~~-2.77,   ~~~~c_{15} &=~~0.878.  \nonumber
\end{eqnarray}
It is interesting that $c_{12} + c_{15} \approx 0$ and that
this results in yet another mass prediction
\begin{eqnarray}
    \Sigma^- - \Sigma^+ &=& (9/5)e +(3/10)(\Lambda + \mu - \Sigma^-)
 \nonumber\\&&~~~~~           - (1/10)(n-p), \nonumber\\
             &=& 7.92 {\rm ~MeV}. \nonumber
\end{eqnarray}
The masses of the eight otherwise unknown baryons also come out
of the theory and are (for this point, in MeV)
\begin{eqnarray}
B_\mu &= 1213.06,~~~~  B_\mu^\prime&= 1107.45, \nonumber\\
B_e &= 1346.61,  ~~~~  B_e^\prime&= 1346.10,   \\
B_1 &= 1120.02,  ~~~~  B_1^\prime &= 1120.02,  \nonumber\\
B_2 &= ~\,941.55,  ~~~~  B_2^\prime &= ~\,941.55.  \nonumber
\end{eqnarray}

\section{Discussion}

     The arrangement shown in Eq.(\ref{E18}) corresponds to case II for
which the quarks have integer charges.  As a matter of fact, the three
points in Fig.~\ref{fig2} lying well within experimental uncertainty
of the measured values all correspond to case II. To conclude
that the ``quarks'' are in fact the proton, antiproton, neutron,
and antineutron (with some admixture of the other baryons and
antibaryons, the $\pi^+$, $\pi^0$, and $\pi^-$
and other mesons thrown in for
good measure) and that the baryons and mesons are self consistent
admixtures of themselves would be quite (even unacceptably)
unconventional. Indeed, the data on the electromagnetic structure
of the nucleons strongly supports the standard model with its
fractionally charged quarks (or partons) (see Ref.~\cite{r3}).
However, one may recall a time
(also 31 years ago) when other
explanations of the electromagnetic structure data were current
and reputable \cite{r9}.

     One would like to have a more accurate experimental value of
$\Xi^-$ - $\Xi^0$.

{\it Acknowledgements.}
JTG was supported by an NRC/NRaD postdoctoral
fellowship through a grant from the ONR. Computational support
was provided by the ACL at Los Alamos National Laboratory.

\appendix{The representation of O(4)}

We give here the
(${3\over2}$,${3\over2}$) representation of O(4) and the connection to O(3),
following Ref.~\cite{r1}.
The finite dimensional irreducible representations of O(4)
are characterized by two spins or iso-spins $S_1$ and $S_2$ and are
denoted by ($S_1$,$S_2$).  The dimensionality of the representation is
(2$S_1$ + 1)(2$S_2$ + 1) corresponding to the fact that the wave
functions forming a basis for such a representation are
distinguished by two quantum numbers $S_{1z}$ and $S_{2z}$,
$-S_1 \le S_{1z} \le S_1$,   $-S_2 \le S_{2z} \le S_2$.

     The main point is that $S_1$ and $S_2$ can be dealt with
separately in all calculations.  By this we mean the following.
The irreducible representations of O(3) are characterized by $S_1$
(everything we say applies equally well to $S_2$).  Let $S_{1z}$
distinguish the wave functions which are a basis for this
2$S_1$+1 dimensional representation of O(3).  The direct product
of these wave functions with themselves can be linearly combined
into bases for irreducible representations of O(3) using the
Clebsch-Gordon coefficients $C$ for O(3) in the well known way:
\begin{eqnarray}
\psi(S_1,S_{1z}) =& \sum
  C(&S_1,S_{1z} ;S_1',S_{1z}';S_1'',S_{1z}'')  \label{A1}\\
    &&\times\psi(S_1',S_{1z}')~\psi(S_1'',S_{1z}''),  \nonumber
\end{eqnarray}
where, as usual,the sum is over $S_{1z}$ and $S_{1z}$' subject to
\begin{eqnarray}
|S_1' - S_1''| \le S_1 \le S_1' + S_1'' ,
&~~&
S_{1z} = S_{1z}' + S_{1z}''      \label{A2}
\end{eqnarray}
The corresponding statements for O(4) are
\begin{eqnarray}
\psi(S_1,&S_{1z}&,S_2,S_{2z}) ~=
       \nonumber\\
& \sum &  C(S_1,S_{1z};S_1',S_{1z}';S_1'',S_{1z}'')
\label{A3}\\
      && \times C(S_2,S_{2z};S_2',S_{2z}';S_2'',S_{2z}'')
       \nonumber\\
       &&\times \psi(S_1',S_{1z}',S_2',S_{2z}')
       ~\psi(S_1'',S_{1z}'',S_2'',S_{2z}'')
\nonumber
\end{eqnarray}
where in addition to Eq.(\ref{A2}) one has the same equations with the
subscript 1 replaced by the subscript 2.
That is, the Clebsch-Gordon coefficients for O(4) are
products of the Clebsch-Gordon coefficients for O(3).

     Now we write down the relevant parts of these results for
the (${3\over2}$,${3\over2}$) representation.  The unnormalized
Clebsch-Gordon coefficients for O(3) give
\FL
\begin{equation}
\left(
\begin{array}{c}\psi(3,0)\\\psi(2,0)\\\psi(1,0)\\\psi(0,0)\end{array}
\right) = \left(
\begin{array}{rrrr}
    1 &  3 &  3 &  1 \\
    1 &  1 & -1 & -1  \\
    3 & -1 & -1 &  3  \\
    1 & -1 &  1 & -1  \\
\end{array}
\right)
\left(
\begin{array}{c}\psi({3\over2},-{3\over2})~\psi({3\over2}, {3\over2})\\
                \psi({3\over2},-{1\over2})~\psi({3\over2}, {1\over2})\\
                \psi({3\over2}, {1\over2})~\psi({3\over2},-{1\over2})\\
                \psi({3\over2}, {3\over2})~\psi({3\over2},-{3\over2})
\end{array} \right)_{.}
\label{A4}
\end{equation}
As before, supposing that $\psi$(0,0) results in all equal masses and
identifying $\bar\psi({3\over2},m)=\pm\psi({3\over2},-m)$, one sees that
the sign of the second and fourth columns of the matrix of Clebsch-Gordon
coefficients must be changed,
\FL
\begin{equation}
\left(
\begin{array}{c}\psi(3,0)\\\psi(2,0)\\\psi(1,0)\\\psi(0,0)\end{array}
\right) = \left(
\begin{array}{rrrr}
    1 & -3 &  3 & -1 \\
    1 & -1 & -1 &  1  \\
    3 &  1 & -1 & -3  \\
    1 &  1 &  1 &  1  \\
\end{array}
\right)
\left(
\begin{array}{c}\bar\psi({3\over2}, {3\over2})~\psi({3\over2}, {3\over2})\\
                \bar\psi({3\over2}, {1\over2})~\psi({3\over2}, {1\over2})\\
                \bar\psi({3\over2},-{1\over2})~\psi({3\over2},-{1\over2})\\
                \bar\psi({3\over2},-{3\over2})~\psi({3\over2},-{3\over2})
\end{array} \right)_{.}
\label{A5}
\end{equation}
The $\psi$(3,0),$\psi$(2,0),$\psi$(1,0),$\psi$(0,0)
are ``mass forms''.  For O(4),
according to Eq.(\ref{A3}), one gets 16 mass forms from the direct
product of the matrix of 1's and 3's in Eq.(\ref{A5}) with itself.
The result is a matrix of 1's, 3's,and 9's.  So that there can be
no confusion, Table~\ref{T3} we give the 16 x 16 matrix $v_{ij}$, with the
quantum numbers to which the elements refer carefully marked.



\begin{table}
\caption{
States, quantum numbers attached to each state,
and electric charge associated with each state in the quark
model. The separation between quartets is of order 100 MeV.
The ``electromagnetic'' separations within each group are of
order 10 MeV.
}\begin{tabular}{lcrrc}

Group   &State   &$S_{1z}$   &$S_{2z}$  &Electric  \\
Name    &Number  &           &          &Charge \\
\hline
        & 1    &${3\over2}$ &${3\over2}$ &3$q_1$ \\
cascade & 2    &            &${1\over2}$ &2$q_1$ + $q_2$ \\
        & 3    &         &$-$${1\over2}$ & $q_1$ +2$q_2$ \\
        & 4    &         &$-$${3\over2}$ &        3$q_2$ \\
\\
        & 5    &${1\over2}$ &${3\over2}$ &2$q_1$         + $q_3$ \\
sigma   & 6    &            &${1\over2}$ &(4$q_1$ +2$q_2$ +2$q_3$ + $q_4$)/3 \\
        & 7    &         &$-$${1\over2}$ &(2$q_1$ +4$q_2$ + $q_3$ +2$q_4$)/3 \\
        & 8    &         &$-$${3\over2}$ &         2$q_2$         + $q_4$  \\
\\
        & 9 &$-$${1\over2}$ &${3\over2}$ &  $q_1$         +2$q_3$ \\
lambda  &10    &            &${1\over2}$ &(2$q_1$ + $q_2$ +4$q_3$ +2$q_4$)/3 \\
        &11    &         &$-$${1\over2}$ &( $q_1$ +2$q_2$ +2$q_3$ +4$q_4$)/3 \\
        &12    &         &$-$${3\over2}$ &          $q_2$         +2$q_4$ \\
\\
        &13 &$-$${3\over2}$ &${3\over2}$ &           3$q_3$ \\
nucleon &14    &            &${1\over2}$ &           2$q_3$ + $q_4$ \\
        &15    &         &$-$${1\over2}$ &            $q_3$ +2$q_4$ \\
        &16    &         &$-$${3\over2}$ &                   3$q_4$ \\
\end{tabular}
\label{T1}
\end{table}

\begin{table}
\caption{
The six possible systems of baryon and quark charges.
}\begin{tabular}{ccrrrrrr}
        &case   &I     &II   &III     &IV    &V    &VI  \\
\hline
      &quark  &\multicolumn{6}{c}{quark charge}           \\
      &$q_1$  &   1/3 &   0 &$-$1/3 &   2/3 &$-$1 &$-$2/3 \\
      &$q_2$  &$-$2/3 &$-$1 &   2/3 &$-$1/3 &   0 &   1/3 \\
      &$q_3$  &   1/3 &   1 &$-$1/3 &   2/3 &   0 &$-$2/3 \\
      &$q_4$  &$-$2/3 &   0 &   2/3 &$-$1/3 &   1 &   1/3 \\
\hline
group   &state &\multicolumn{6}{c}{baryon charge}        \\
          & 1  &   1  &   0  &$-$1  &   2   &$-$3  &$-$2 \\
cascade   & 2  &   0  &$-$1  &   0  &   1   &$-$2  &$-$1 \\
group     & 3  &$-$1  &$-$2  &   1  &   0   &$-$1  &   0 \\
          & 4  &$-$2  &$-$3  &   2  &$-$1   &   0  &   1 \\
\\
          & 5  &   1  &   1  &$-$1  &   2   &$-$2  &$-$2 \\
sigma     & 6  &   0  &   0  &   0  &   1   &$-$1  &$-$1 \\
group     & 7  &$-$1  &$-$1  &   1  &   0   &   0  &   0 \\
          & 8  &$-$2  &$-$2  &   2  &$-$1   &   1  &   1 \\
\\
          & 9  &   1  &   2  &$-$1  &   2   &$-$1  &$-$2 \\
lambda    &10  &   0  &   1  &   0  &   1   &   0  &$-$1 \\
group     &11  &$-$1  &   0  &   1  &   0   &   1  &   0 \\
          &12  &$-$2  &$-$1  &   2  &$-$1   &   2  &   1 \\
\\
          &13  &   1  &   3  &$-$1  &   2   &   0  &$-$2 \\
nucleon   &14  &   0  &   2  &   0  &   1   &   1  &$-$1 \\
group     &15  &$-$1  &   1  &   1  &   0   &   2  &   0 \\
          &16  &$-$2  &   0  &   2  &$-$1   &   3  &   1 \\
\end{tabular}
\label{T2}
\end{table}

\widetext
\begin{table}
\caption{
The Matrix (${\bf v}$) for O(4)
}\begin{tabular}{rrrrrrrrrrrrrrrrrr}
&&\multicolumn{16}{c}{mass form number} \\
  $S_{1z}$ &  $S_2$   &1 &2 &3 &4 &5 &6 &7 &8 &9&10&11&12&13&14&15&16 \\
\hline
   ${3\over2}$ &   ${3\over2}$
    &1& 3& 1& 1& 3& 9& 3& 3& 1& 3& 1& 1& 1& 3& 1& 1      \\
   ${3\over2}$ &   ${1\over2}$
    &1& 3& 1& 1& 1& 3& 1& 1&$-$1&$-$3&$-$1&$-$1&$-$3&$-$9&$-$3&$-$3 \\
   ${3\over2}$ &$-$${1\over2}$
    &1& 3& 1& 1&$-$1&$-$3&$-$1&$-$1&$-$1&$-$3&$-$1&$-$1& 3& 9& 3& 3 \\
   ${3\over2}$ &$-$${3\over2}$
    &1& 3& 1& 1&$-$3&$-$9&$-$3&$-$3& 1& 3& 1& 1&$-$1&$-$3&$-$1&$-$1 \\
   ${1\over2}$ &   ${3\over2}$
    &1& 1&$-$1&$-$3& 3& 3&$-$3&$-$9& 1& 1&$-$1&$-$3& 1& 1&$-$1&$-$3 \\
   ${1\over2}$ &   ${1\over2}$
    &1& 1&$-$1&$-$3& 1& 1&$-$1&$-$3&$-$1&$-$1& 1& 3&$-$3&$-$3& 3& 9 \\
   ${1\over2}$ &$-$${1\over2}$
    &1& 1&$-$1&$-$3&$-$1&$-$1& 1& 3&$-$1&$-$1& 1& 3& 3& 3&$-$3&$-$9 \\
   ${1\over2}$ &$-$${3\over2}$
    &1& 1&$-$1&$-$3&$-$3&$-$3& 3& 9& 1& 1&$-$1&$-$3&$-$1&$-$1& 1& 3 \\
$-$${1\over2}$ &   ${3\over2}$
    &1&$-$1&$-$1& 3& 3&$-$3&$-$3& 9& 1&$-$1&$-$1& 3& 1&$-$1&$-$1& 3 \\
$-$${1\over2}$ &   ${1\over2}$
    &1&$-$1&$-$1& 3& 1&$-$1&$-$1& 3&$-$1& 1& 1&$-$3&$-$3& 3& 3&$-$9 \\
$-$${1\over2}$ &$-$${1\over2}$
    &1&$-$1&$-$1& 3&$-$1& 1& 1&$-$3&$-$1& 1& 1&$-$3& 3&$-$3&$-$3& 9 \\
$-$${1\over2}$ &$-$${3\over2}$
    &1&$-$1&$-$1& 3&$-$3& 3& 3&$-$9& 1&$-$1&$-$1& 3&$-$1& 1& 1&$-$3 \\
$-$${3\over2}$ &   ${3\over2}$
    &1&$-$3& 1&$-$1& 3&$-$9& 3&$-$3& 1&$-$3& 1&$-$1& 1&$-$3& 1&$-$1 \\
$-$${3\over2}$ &   ${1\over2}$
    &1&$-$3& 1&$-$1& 1&$-$3& 1&$-$1&$-$1& 3&$-$1& 1&$-$3& 9&$-$3& 3 \\
$-$${3\over2}$ &$-$${1\over2}$
    &1&$-$3& 1&$-$1&$-$1& 3&$-$1& 1&$-$1& 3&$-$1& 1& 3&$-$9& 3&$-$3 \\
$-$${3\over2}$ &$-$${3\over2}$
    &1&$-$3& 1&$-$1&$-$3& 9&$-$3& 3& 1&$-$3& 1&$-$1&$-$1& 3&$-$1& 1 \\
\end{tabular}
\label{T3}
\end{table}
\narrowtext

\figure{\label{fig1}
A wide view of the 99,792 mass predictions for $\Xi^- - \Xi^0$ and
$\Sigma^- - \Sigma^0$. Most of the mass predictions lie on ``avenues''.
A closer view of the mass predictions in the box, which lie in the
vicinity of the experimental values, is given in Fig.~2.        }

\figure{\label{fig2}
The mass predictions in the vicinity of the
experimental values. Here the box denotes the
experimental uncertainty \cite{r5} in the values of
$\Xi^- - \Xi^0$ and $\Sigma^- - \Sigma^0$.
The empty space to the right of the box yields the prediction
that more precise values must satisfy $\Xi^- - \Xi^0 < 6.1$ MeV,
in disagreement with the prediction of the Coleman-Glashow \cite{r4}
formula,  $\Xi^- - \Xi^0 = 6.78 \pm 0.08$ MeV. Currently the
experimental uncertainty in the mass data is too large to
decide between the standard model \cite{r3} and the
highly unconventional theory presented in this paper.       }
\end{narrowtext}

\begin{references}
\bibitem{r3} See F.E.~Close,
{\it An Introduction to Quarks and Partons}
(Academic, New York, 1979).
\bibitem{r4} Ref.~\cite{r3}, p.403.
\bibitem{r5} Experimental values are taken from
{\it Phys.~Rev.~D}{\bf 45}, Part II (1992)
({\it Review of Particle Properties}).
\bibitem{r0} We stress again that we present here
a mathematical model which fits experimental mass
data yet disagrees with the standard model in the hope to stimulate
an effort to improve the accuracy of the $\Xi^-$ - $\Xi^0$.
Physically, the model is far fetched for several reasons.
First, the work may appear to be an effort toward grand unification.
It is not -- the $B$ baryons have masses
whose differences happen to be lepton masses,
but the leptons are not otherwise included.
The difficulties in constructing grand unification with O(4)
are well known
[L. O'Raighfeartaigh, {\it Group Structure of Guage Theories},
(Cambridge University Press, Cambridge, 1986), Sec. 5.4].
Even if this work were an effort toward grand unification it would
fail badly because the $\tau$ lepton and its neutrino, as well as
other important particles such as the $\Omega^-$, are not included.
\bibitem{r6} The one mass in the list which is
unusual is $\Lambda + \mu - \Sigma^-$.
This may be seen to be an ``electromagnetic'' mass by writing it in the
form $(\Lambda + B_\mu^\prime) - (\Sigma^- - B_\mu)$.
\bibitem{r7} The $m_j^\prime$ involve large cancellations such as $n-p$.
These cancellations also appear in the Coleman-Glashow formula and are
not the cancellations to which we refer. We refer to cancellations
between two different $m_j^\prime$.
\bibitem{r8} In our view, only a prediction which is consistent with
all 99,792 points is worthy of note.
\bibitem{r9} S.D. Drell and F. Zachariasen,
{\it Electromagnetic Structure of the Nucleons}
(Oxford University Press, London, 1961).
\bibitem{r1} I.M.~Gel`fand, R.A. Minlos, and Z.Ya. Shapiro,
{\it Representations of the Rotation and Lorentz Groups
and their Applications} (MacMillan Company, New York, 1963).
\end{references}
\end{document}